\documentclass[11pt]{article}

\usepackage[margin=1.05in]{geometry}
\usepackage{amsmath,amssymb}
\usepackage{booktabs}
\usepackage{graphicx}
\usepackage[font=small,labelfont=bf]{caption}
\usepackage[round]{natbib}
\usepackage{fontspec}
\usepackage{microtype}
\usepackage{xcolor}
\definecolor{linkcol}{RGB}{20,70,140}
\usepackage[colorlinks=true,allcolors=linkcol]{hyperref}

\newfontfamily\cjkfont{NotoSerifCJKsc-subset.otf}[Path=./]
\newcommand{\zh}[1]{{\cjkfont #1}}

\newcommand{\Om}{\Omega}
\title{\bfseries Zipf's Law of Abbreviation in a Logographic Script:\\
Coding-Theoretic Bounds on Chinese Character Stroke Counts}

\author{Mustafa Ergen\thanks{Correspondence: \texttt{mustafa.ergen@cloudoneai.ai}.}}

\date{August 2026}

\begin{document}
\maketitle

\begin{abstract}
\noindent
Zipf's law of abbreviation --- the tendency of frequent forms to be short --- is one of the
best-supported quantitative regularities in language, and recent work has moved from
demonstrating its existence to measuring \emph{how far} lexicons are compressed relative to
principled baselines. That programme has so far been carried out on word lengths in alphabetic
and syllabic scripts. We transfer it to a logographic script, taking the stroke as the unit of
articulatory cost and the Chinese character as the coded form. Combining a stroke-order database
covering all 20{,}902 characters of the CJK basic block with two independent frequency corpora
(258.9M and 193.3M tokens), we find that the mean character \emph{type} costs 12.71 strokes but
the mean character \emph{token} in running text only 7.22. Using the dually normalised optimality
score of \citet{petrini2026optimality}, the simplified Chinese inventory reaches
$\Om = 0.668$, with the replication corpus at $0.609$ --- inside and just below the 62--67\% band
those authors report for word lengths across 20 languages and 8 scripts, which suggests a
compression ceiling largely independent of script type and cost unit. A logographic script additionally makes
\emph{absolute} coding bounds computable, because strokes are drawn from a closed five-element
taxonomy: the exact 5-ary Huffman optimum is 4.34 strokes and the entropy bound is 4.28 strokes,
so the observed system is $1.66\times$ above optimal coding. We show that this residual gap is not
slack but structure. The Kraft sum $\sum_i 5^{-\ell_i}$ equals 2.05 on the frequency list and
5.03 on the full inventory, so stroke strings are provably not uniquely decodable in one
dimension; characters are disambiguated by the two-dimensional arrangement of strokes rather than
by their sequence, and the forgone compression buys componential (semantic--phonetic)
transparency. Finally, treating the mid-twentieth-century simplification reform as a controlled
compression event, we find it raised optimality from $\Om = 0.555$ to $\Om = 0.668$, with savings
concentrated in the most frequent 1{,}000 characters --- precisely the intervention an optimal
coder would make.
\end{abstract}

\section{Introduction}

Communication systems that reuse a finite repertoire of forms face a recurring pressure: forms
that are used often should be cheap to produce. \citet{zipf1949} elevated this observation to a
principle of least effort, and the resulting \emph{law of abbreviation} --- an inverse relation
between the frequency of a form and its magnitude --- has since been documented in 986 languages
\citep{bentz2016universal}, in speech duration as well as orthographic length
\citep{petrini2023direct}, and in the vocal repertoires of other species. It is also exactly what
optimal coding predicts: \citet{shannon1948} bounds the mean code length of a source by its
entropy, and \citet{huffman1952} gives a constructive procedure whose codeword lengths decrease
with symbol probability.

The interesting scientific question is therefore no longer \emph{whether} lexicons obey the law,
but \emph{how close to optimal} they are. Answering that requires baselines. Recent work has
supplied them: \citet{petrini2023direct} derive a random baseline showing that word lengths sit
systematically below chance, and \citet{petrini2026optimality} define dually normalised optimality
scores --- normalised against both a random and a minimum baseline --- and report that word
lengths are optimised to roughly 62--67\% on average across 20 languages from 9 families written
in 8 scripts, and to about 65\% when measured in speech time.

That programme has been conducted almost entirely on \emph{word} lengths in phonographic scripts,
where the unit of cost is a letter or a unit of time. This paper asks what happens in a
logographic script, where the coded form is a character and the natural unit of production cost is
the brush or pen stroke. Chinese is an unusually favourable test case for three reasons. First,
the character is a strong functional analogue of the word for readers of Chinese
\citep{deng2014rank}. Second, complete stroke-order data exist, so the cost of every character is
known exactly rather than estimated. Third --- and this is what makes the logographic case
strictly more informative than the alphabetic one --- strokes are drawn from a \emph{closed
five-element taxonomy}. That closure turns a relative question into an absolute one: we can
compute the exact optimum of a 5-ary Huffman code over the same frequency distribution, evaluate
the Shannon bound in the same units, and --- most diagnostically --- test the Kraft inequality to
ask whether stroke strings could function as a uniquely decodable code at all.

Our contributions are:

\begin{enumerate}
\itemsep2pt
\item We transfer the dually normalised optimality framework to a logographic script with the
stroke as cost unit, and report $\Om = 0.668$ for simplified Chinese, replicated at $\Om = 0.609$
on an independent corpus of a different register (Section~\ref{sec:opt}). Both values fall in or
adjacent to the band reported for word lengths across unrelated scripts, which we read as evidence
for a compression ceiling that is not an artefact of any one script type or cost unit.
\item We compute absolute coding-theoretic references that are unavailable in the word-length
setting --- the exact 5-ary Huffman optimum, the entropy bound in strokes, and order-0 and
order-1 stroke-type entropy rates (Section~\ref{sec:abs}).
\item We introduce the Kraft sum as a structural diagnostic for writing systems and show that it
explains the residual gap: at 2.05 (frequency list) and 5.03 (full inventory) it decisively
exceeds unity, so the script cannot be a uniquely decodable one-dimensional code, and the
compression it forgoes is what pays for spatial and componential structure
(Section~\ref{sec:kraft}).
\item We quantify the twentieth-century simplification reform as a deliberate compression event
that moved $\Om$ from 0.555 to 0.668, with the saving concentrated in high-frequency characters
(Section~\ref{sec:reform}).
\end{enumerate}

We are explicit about what is \emph{not} new. That Chinese characters obey the law of abbreviation
in stroke counts is expected and has been observed before; our negative correlations
($\rho \approx -0.55$) are a replication, not a discovery. The contribution is the measurement
against principled bounds, and the identification of the mechanism that caps it.

\section{Related work}

\paragraph{Law of abbreviation and compression.}
\citet{zipf1949} is the origin; \citet{bentz2016universal} establish near-universality across 986
languages; \citet{ferrer2015compression} show formally that minimising a mean energetic cost
function over form probability and magnitude and a negative probability--magnitude correlation are
intimately related, giving the law a coding-theoretic derivation.
\citet{petrini2023direct} extend the evidence to speech duration in 46 languages and, crucially,
introduce a random baseline that converts the qualitative law into a quantitative deficit.
\citet{petrini2026optimality} then define dually normalised optimality scores and estimate degrees
of optimality per language; we adopt their $\Om$ directly so that our numbers are comparable to
theirs.

\paragraph{Chinese characters as a statistical system.}
\citet{deng2014rank} show that rank--frequency relations for Chinese characters follow a
two-layer structure --- Zipfian for frequent characters, exponential-like in the tail --- and
argue that characters play for Chinese writers a role comparable to words in alphabetic systems;
this licenses treating the character as the coded form. \citet{junda2004} provides the
character-frequency resource used here. Stroke count is the standard operationalisation of
character visual complexity in the psycholinguistic literature, where it predicts naming latency
and fixation duration, although recognition appears to be limited by overall complexity rather
than by stroke count alone --- a caveat we return to in Section~\ref{sec:limits}.

\paragraph{Information theory applied to orthography.}
Information-theoretic measures have been used to characterise orthography--phonology mappings and
to compare the transparency of writing systems. To our knowledge, however, the Kraft inequality
has not previously been used as a diagnostic of a writing system's coding architecture, and no
prior work computes a Huffman optimum in the script's own native cost unit for comparison with the
observed inventory.

\section{Data}

\paragraph{Stroke data.}
We use a stroke-order database covering the 20{,}902 characters of the CJK Unified Ideographs
basic block (U+4E00--U+9FA5), distributed with the \texttt{chinese-stroke-sorting} package. Each
character is represented as a string over the five-element stroke taxonomy of the PRC standard:
\zh{横} \emph{héng} (horizontal, coded \texttt{1}), \zh{竖} \emph{shù} (vertical, \texttt{2}),
\zh{撇} \emph{piě} (left-falling, \texttt{3}), \zh{点} \emph{diǎn} (dot / right-falling,
\texttt{4}) and \zh{折} \emph{zhé} (turning, \texttt{5}). For example
\zh{的} $\rightarrow$ \texttt{32511354}. The length of the string is the stroke count, so the
resource gives both the cost of each character and the identity of each of its strokes. Mean
inventory cost over all 20{,}902 characters is 12.84 strokes.

We validated the database against 32 characters whose standard stroke counts we could check
independently (the list and the check are in \texttt{scripts/validate\_strokes.py} in the
repository); 31 matched exactly. The single discrepancy (\zh{體} coded as 22 rather than 23)
reflects the PRC convention of counting the \zh{骨} component as 9 strokes rather than the 10 used
in Taiwan, confirming that the resource follows the PRC standard consistently.

\paragraph{Frequency data.}
Our primary corpus is the character-frequency list of \citet{junda2004}, comprising
258{,}851{,}453 tokens of modern written Chinese; 11{,}991 of its types have stroke data.
As an independent replication we use the character-frequency list derived from the Leiden Weibo
Corpus \citep{vanesch2012}, 193{,}343{,}550 tokens of Chinese social-media text, of which 8{,}911
types have stroke data. The two corpora differ in register (edited written prose versus informal
microblog text) and were collected independently. Their character log-probabilities agree closely
on the 8{,}911 shared types (Spearman $\rho = 0.974$), but they are not interchangeable: the
Leiden inventory is smaller and its type inventory is less costly on average (11.73 versus 12.71
strokes), which matters for the baselines below.

\paragraph{Simplified--traditional mapping.}
To evaluate the simplification reform we use the single-character entries of CC-CEDICT, which
yield traditional counterparts for 10{,}799 simplified characters. Of the 11{,}991 characters in
the primary frequency list, 2{,}415 have a traditional counterpart with a different stroke count
(2{,}479 across the full 20{,}902-character inventory).
Characters with no mapping are assigned their own stroke count in the pre-reform condition, so our
estimate of the reform's effect is conservative.

\section{Methods}

\subsection{Setup}
Let the character types be indexed $i = 1, \dots, n$ with token probabilities $p_i$ (estimated by
relative frequency) and stroke counts $\ell_i$. The quantity of interest is the mean cost per
token,
\begin{equation}
L \;=\; \sum_{i=1}^{n} p_i \ell_i ,
\end{equation}
which we contrast with the unweighted inventory mean $\frac{1}{n}\sum_i \ell_i$.

\subsection{Baselines and the optimality score}
Following \citet{petrini2026optimality} we use three baselines, all of which hold the
\emph{multiset} of stroke counts and the frequency distribution fixed and vary only their pairing.
The random baseline is the expectation of $L$ under a uniformly random assignment,
\begin{equation}
L_r \;=\; \mathbb{E}_\pi\!\left[\textstyle\sum_i p_i \ell_{\pi(i)}\right] \;=\;
\frac{1}{n}\sum_{i=1}^{n}\ell_i ,
\end{equation}
i.e.\ the inventory mean. The minimum baseline $L_{\min}$ pairs the ascending sequence of stroke
counts with the descending sequence of probabilities, and the maximum baseline $L_{\max}$ reverses
this; both are optimal by the rearrangement inequality. The dually normalised optimality score is
\begin{equation}
\Om \;=\; \frac{L_r - L}{L_r - L_{\min}} ,
\label{eq:omega}
\end{equation}
so $\Om = 0$ at chance and $\Om = 1$ when the existing inventory of shapes is assigned to
meanings as efficiently as possible. Because $\Om$ is normalised against both baselines it is
comparable across languages, scripts and cost units, which is what lets us place our figure
alongside published word-length results.

\subsection{Absolute bounds}
The stroke taxonomy has exactly $D = 5$ elements, so we can additionally compute references that
do not depend on the existing inventory. Let $H = -\sum_i p_i \log_2 p_i$. We compute (i) the
exact $D$-ary Huffman code length \citep{huffman1952} over $(p_i)$, padding the alphabet with
zero-probability symbols so that $n \equiv 1 \pmod{D-1}$, giving $L_{\mathrm{Huff}}$; (ii) the
Shannon bound $H/\log_2 D$, the mean number of strokes needed if every stroke carried
$\log_2 5$ bits; and (iii) tighter bounds $H/H_0$ and $H/H_1$, where $H_0$ and $H_1$ are the
token-weighted order-0 and order-1 entropy rates of the stroke sequences themselves, which
account for the fact that stroke types are neither equiprobable nor independent.

\subsection{Kraft sum}
A uniquely decodable $D$-ary code must satisfy the Kraft--McMillan inequality
\citep{kraft1949,mcmillan1956}
\begin{equation}
\sum_{i=1}^{n} D^{-\ell_i} \;\le\; 1 .
\label{eq:kraft}
\end{equation}
We evaluate the left-hand side with $D = 5$ over both the frequency list and the full 20{,}902
character inventory. Unlike the optimality score, this is not a measure of how well the system is
tuned; it is a test of whether the system can be a one-dimensional code at all.

\subsection{Inference}
We assess the observed $L$ against the random baseline by permutation: 2{,}000 random
reassignments of stroke counts to characters give a null distribution of $L$, from which we report
a $z$ score (500 permutations for the reform conditions and 300 for the truncated inventories,
where $z$ is reported only for orientation). We report Spearman correlations between $\log$ frequency and stroke count, and we
repeat the whole pipeline on the replication corpus, on the pre-reform stroke assignment, and on
truncated inventories (Section~\ref{sec:robust}).

\section{Results}

\subsection{The law of abbreviation in strokes}

\begin{figure}[t]
\centering
\includegraphics[width=\linewidth]{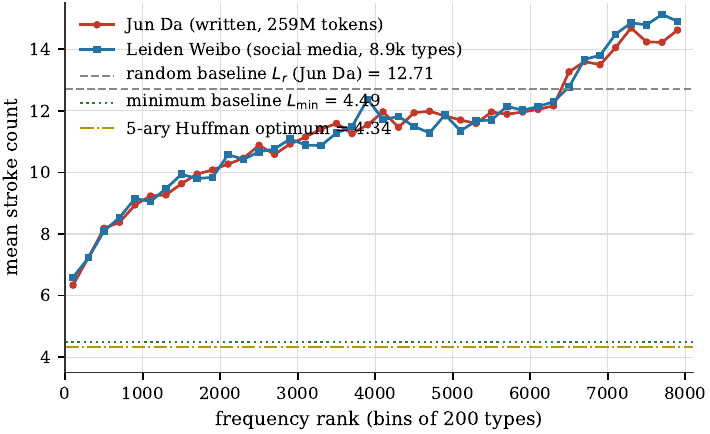}
\caption{Mean stroke count by frequency rank, in bins of 200 types, for both corpora. The relation
is close to monotone across four orders of magnitude of frequency \emph{rank} and replicates
across register.
Horizontal lines give the random baseline, the minimum baseline and the 5-ary Huffman optimum for
the primary corpus.}
\label{fig:rank}
\end{figure}

Stroke count increases almost monotonically with frequency rank (Figure~\ref{fig:rank}). In the
primary corpus the 100 most frequent characters average 6.04 strokes and cover 38.9\% of all
tokens, while the 3{,}991 rarest average 15.53 strokes and cover under 0.01\% of
tokens (Table~\ref{tab:bands}).
Spearman $\rho$ between $\log$ frequency and stroke count is $-0.545$ (primary) and $-0.495$
(replication). Reading a median text is correspondingly cheap: 179 types cover half of all tokens
at a mean cost of 6.30 strokes, and 1{,}265 types cover 90\% at 8.13 strokes.

\begin{table}[t]
\centering
\small
\caption{Stroke cost by frequency band, primary corpus ($n = 11{,}991$).}
\label{tab:bands}
\begin{tabular}{lrrr}
\toprule
Rank band & Types & Mean strokes & Token coverage \\
\midrule
1--100 & 100 & 6.04 & 38.91\% \\
101--500 & 400 & 7.32 & 33.19\% \\
501--1{,}000 & 500 & 8.57 & 14.08\% \\
1{,}001--2{,}000 & 1{,}000 & 9.63 & 9.38\% \\
2{,}001--3{,}000 & 1{,}000 & 10.62 & 2.77\% \\
3{,}001--5{,}000 & 2{,}000 & 11.61 & 1.41\% \\
5{,}001--8{,}000 & 3{,}000 & 13.03 & 0.26\% \\
8{,}001--11{,}991 & 3{,}991 & 15.53 & $<$0.01\% \\
\bottomrule
\end{tabular}
\end{table}

The aggregate relation is strong but individually noisy: regressing stroke count on
$\log_5(\text{rank})$ gives $\ell = -6.94 + 3.77\log_5(\text{rank})$ with $R^2 = 0.26$. Frequency
constrains the average cost of a rank band tightly while leaving the cost of any individual
character largely free --- a pattern consistent with a diffuse selective pressure rather than
per-character design.

\begin{figure}[t]
\centering
\includegraphics[width=\linewidth]{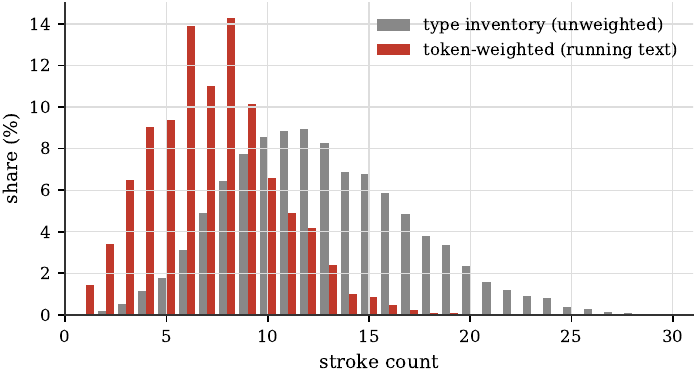}
\caption{Stroke-count distributions over the type inventory (unweighted) and over tokens in
running text (frequency-weighted), primary corpus.}
\label{fig:dist}
\end{figure}

The type/token contrast makes the same point in a single pair of numbers. The mean character in
the inventory costs 12.71 strokes; the mean character \emph{token} in running text costs 7.22
strokes (Figure~\ref{fig:dist}). The modal bins of the inventory distribution are 10--13 strokes; those of the
token distribution are 5--9.

\subsection{Degree of optimality}
\label{sec:opt}

Table~\ref{tab:main} collects the baselines and scores. With $L = 7.218$, $L_r = 12.706$ and
$L_{\min} = 4.491$, Equation~\ref{eq:omega} gives $\Om = 0.668$. The observed value lies
21.8 null standard deviations below the permutation mean, so the arrangement is not attributable
to chance under any reasonable tolerance. The replication corpus yields $\Om = 0.609$
($z = -16.7$); the difference between the two is driven mainly by their different type
inventories rather than by disagreement about frequencies, which are near-identical
($\rho = 0.974$ on shared types).

\begin{table}[t]
\centering
\small
\caption{Baselines, optimality and absolute bounds. All costs in strokes per character token.}
\label{tab:main}
\begin{tabular}{lrr}
\toprule
& Jun Da (written) & Leiden Weibo (social) \\
Types / tokens & 11{,}991 / 258.9M & 8{,}911 / 193.3M \\
\midrule
Maximum baseline $L_{\max}$ & 24.169 & 22.699 \\
Random baseline $L_r$ & 12.706 & 11.730 \\
\textbf{Observed} $L$ & \textbf{7.218} & \textbf{7.203} \\
Minimum baseline $L_{\min}$ & 4.491 & 4.293 \\
5-ary Huffman optimum & 4.340 & 4.222 \\
Entropy bound $H/\log_2 5$ & 4.279 & 4.161 \\
\midrule
Optimality $\Om$ & \textbf{0.668} & \textbf{0.609} \\
Permutation $z$ & $-21.8$ & $-16.7$ \\
Spearman $\rho(\log f, \ell)$ & $-0.545$ & $-0.495$ \\
$H$ (bits/character) & 9.936 & 9.661 \\
Kraft sum $\sum 5^{-\ell_i}$ & 2.053 & 2.047 \\
\bottomrule
\end{tabular}
\end{table}

The comparison we consider most informative is external. \citet{petrini2026optimality} report mean
optimality of 62--67\% for word lengths measured in characters across 20 languages of 9 families
in 8 scripts, and about 65\% for word durations in speech across 46 languages. Our logographic,
stroke-based estimates of 0.668 and 0.609 fall inside and just below that band. Given how little
the two settings share --- different linguistic unit (character versus word), different cost unit
(stroke versus letter or millisecond), different script family --- the convergence suggests that
the two-thirds figure is not an artefact of alphabetic writing but reflects a general ceiling on
how far a communicative code can be compressed while remaining usable. We stress that this is a
suggestive alignment across independently conducted studies, not a controlled comparison.

\subsection{Absolute bounds}
\label{sec:abs}

\begin{figure}[t]
\centering
\includegraphics[width=\linewidth]{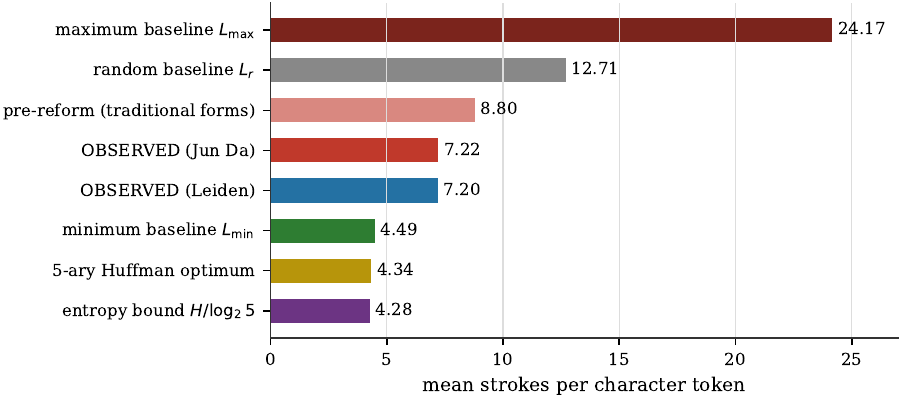}
\caption{The optimality landscape for the primary corpus, with the observed value of the
replication corpus and the pre-reform (traditional-form) condition for comparison. Bars are mean
strokes per character token.}
\label{fig:opt}
\end{figure}

Because the stroke alphabet is closed, we can go further than relative baselines
(Figure~\ref{fig:opt}). The exact 5-ary Huffman optimum over the primary frequency distribution is
4.340 strokes per token and the entropy bound is 4.279, so the observed 7.218 is $1.66\times$
optimal coding. Notably, the minimum baseline $L_{\min} = 4.491$ sits only 3.5\% above the Huffman
optimum: the existing stock of character shapes is, as a multiset of lengths, almost exactly the
right stock for this frequency distribution. What the system does not do is assign those shapes to
meanings optimally.

Stroke types are not equiprobable. Token-weighted, \zh{横} accounts for 29.8\% of all strokes
written, followed by \zh{竖} (18.9\%), \zh{折} (17.7\%), \zh{点} (17.0\%) and \zh{撇} (16.6\%),
giving an order-0 entropy rate of $H_0 = 2.281$ bits per stroke against a maximum of
$\log_2 5 = 2.322$; adding first-order dependencies between consecutive strokes lowers this to
$H_1 = 2.102$. Even the most permissive of these --- charging characters only $H_1$ bits per
stroke --- puts the bound at $H/H_1 = 4.73$ strokes, still well below the observed 7.22. In the
other direction, the observed system conveys $H/L = 1.38$ bits per stroke, about 59\% of the
$\log_2 5$ ceiling.

\subsection{Why the gap does not close}
\label{sec:kraft}

The natural inference from a $1.66\times$ gap would be that the script is simply not very well
compressed. The Kraft sum shows that this inference is wrong in an interesting way. Evaluating
Equation~\ref{eq:kraft} with $D=5$ gives 2.053 over the 11{,}991 characters of the frequency list
and 5.031 over the full 20{,}902-character inventory. Both decisively exceed 1, and by
Kraft--McMillan no uniquely decodable 5-ary code can have these length profiles. Stroke strings
are therefore not merely a suboptimal code --- they are not a code in the required sense at all.
Two observations make this concrete. First, prefixes are everywhere: \zh{一} is the single stroke
\texttt{1}, and 6{,}197 of the 20{,}902 characters begin with a horizontal stroke, so nearly a
third of the inventory has \zh{一} as a prefix. Second, and more decisively, the mapping is not
even non-singular: distinct characters share identical stroke strings. \zh{人} `person',
\zh{入} `enter' and \zh{八} `eight' are all \texttt{34}; \zh{士} `scholar' and \zh{土} `earth'
are both \texttt{121}, differing only in which horizontal stroke is longer; \zh{日} `sun' and
\zh{曰} `to say' are both \texttt{2511}; \zh{未} and \zh{末} are both \texttt{11234}. Across the
full inventory, 878 distinct stroke strings are shared by two or more characters, involving 1{,}943
characters in total, with one string (\texttt{354}) shared by nine.

What disambiguates Chinese characters is the two-dimensional arrangement of strokes, not their
linear order. This is the correct explanation for the residual gap, and it reframes the gap as a
price rather than a failure. Spending strokes redundantly is what allows characters to be built
from recurring components in fixed spatial slots, which in turn is what makes the system partially
transparent: a reader who knows \zh{水} `water' can guess that \zh{河}, \zh{江}, \zh{湖} and
\zh{海} concern water, and one who knows \zh{言} `speech' can guess the domain of \zh{语},
\zh{说}, \zh{话} and \zh{谈}. An optimal Huffman code makes such inference impossible by
construction, since codeword neighbourhoods carry no meaning. On this reading the script is not
solving the compression problem badly; it is solving a different problem --- compression subject
to learnability and componential transparency --- and the Kraft sum measures the constraint.

\subsection{The simplification reform as a compression event}
\label{sec:reform}

The mid-twentieth-century simplification of Chinese characters offers something rare: a
deliberate, documented intervention on the cost side of the system, with the frequency
distribution approximately held fixed. Substituting traditional stroke counts for the same 11{,}991
characters and re-running the pipeline gives a pre-reform mean of 8.795 strokes per token and
$\Om = 0.555$; the simplified inventory gives 7.218 and $\Om = 0.668$
(Table~\ref{tab:reform}). The reform thus removed 1.58 strokes per token, a 17.9\% reduction, and
raised measured optimality by 11.3 percentage points.

\begin{table}[t]
\centering
\small
\caption{The simplification reform, primary frequency distribution held fixed. Traditional stroke
counts from CC-CEDICT single-character mappings; unmapped characters retain their own count, so
the pre-reform figures are conservative.}
\label{tab:reform}
\begin{tabular}{lrr}
\toprule
& Traditional forms & Simplified forms \\
\midrule
Observed $L$ & 8.795 & 7.218 \\
Random baseline $L_r$ & 13.733 & 12.706 \\
Minimum baseline $L_{\min}$ & 4.842 & 4.491 \\
Optimality $\Om$ & 0.555 & \textbf{0.668} \\
Spearman $\rho(\log f, \ell)$ & $-0.372$ & $-0.545$ \\
\bottomrule
\end{tabular}
\end{table}

\begin{figure}[t]
\centering
\includegraphics[width=\linewidth]{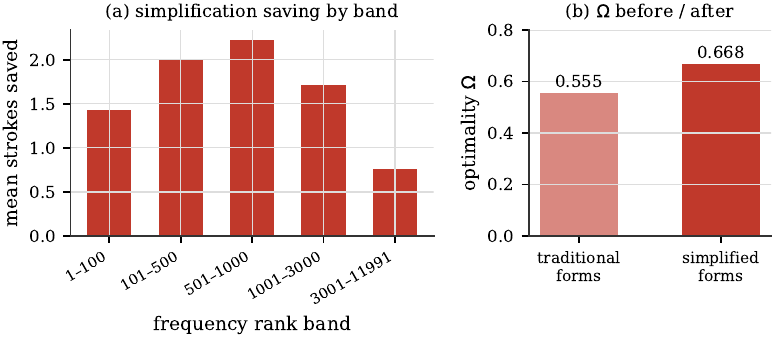}
\caption{(a) Mean strokes saved by simplification, by frequency rank band. (b) Optimality $\Om$
under traditional and simplified forms with the frequency distribution held fixed.}
\label{fig:reform}
\end{figure}

More telling than the magnitude is the targeting. Mean strokes saved per character is 1.43 in
ranks 1--100, 2.00 in 101--500 and 2.23 in 501--1{,}000, but only 0.76 in the tail beyond rank
3{,}000, where just 15.1\% of characters were altered at all against 26--38\% in the frequent
bands (Figure~\ref{fig:reform}). This is the shape of intervention an optimal coder would choose:
shorten the high-probability symbols and leave the tail alone. Two qualifications matter. First,
$\Om = 0.555$ for traditional forms shows that most of the compression predates the reform and was
achieved by ordinary use over a much longer period; the reform sharpened an existing pattern
rather than creating it. Second, our pre-reform condition holds modern frequencies fixed, so it
measures the counterfactual cost of writing \emph{today's} texts in traditional forms, not the
historical state of the system.

\subsection{Robustness}
\label{sec:robust}

The result is not driven by outliers. \zh{的} alone accounts for 3.21\% of all tokens at 8 strokes
and is the single costliest character relative to its ideal length; removing it \emph{raises}
$\Om$ to 0.679. Removing the top 10 gives 0.664 and the top 100 gives 0.635, all close to the
full-inventory value.

$\Om$ is, however, sensitive to how much of the tail is included, and this deserves emphasis
because it constrains cross-study comparison. Restricting the analysis to the 1{,}000, 3{,}000 and
6{,}000 most frequent characters gives $\Om = 0.377$, $0.492$ and $0.559$ respectively. The reason
is structural rather than substantive: truncating the inventory compresses the gap between $L_r$
and $L_{\min}$, so the same absolute deviation registers as a smaller fraction. Optimality scores
of this family are therefore only comparable between studies that use comparably deep inventories
--- a point that applies equally to published word-length estimates and that we flag as a caveat
on our own external comparison in Section~\ref{sec:opt}.

\section{Discussion}

Three findings seem worth separating.

The first is a replication with a twist. Chinese characters obey the law of abbreviation in
strokes, as expected; what is new is that when the law is measured against principled baselines,
the logographic answer lands where the alphabetic answers land. If the two-thirds figure survives
in further scripts and cost units, it becomes a target for explanation in its own right: what
property of communication systems stops compression at roughly two-thirds of the achievable
range, rather than at 40\% or 95\%?

The second is that the Chinese inventory is better at supplying the right \emph{distribution} of
costs than at \emph{allocating} them. The minimum baseline is within 3.5\% of the Huffman optimum,
so the stock of shapes is nearly ideal for this frequency profile; the entire 2.7-stroke shortfall
between $L$ and $L_{\min}$ is misallocation --- shapes of the right lengths attached to the wrong
meanings. This is what a system shaped by incremental, local pressure should look like: the
distribution of form complexity can be sculpted by aggregate wear, but globally reassigning forms
to meanings is not an operation that writing communities can perform.

The third is methodological. The Kraft sum is cheap to compute for any writing system with a
finite grapheme inventory, and it answers a question that correlations and optimality scores
cannot: whether the system is even the kind of object that optimal-coding bounds apply to. For
Chinese the answer is no, and that negative answer explains the residual gap better than any
appeal to inefficiency. We would expect the same diagnostic to be informative for other scripts
where graphemes are composed spatially rather than concatenated --- Korean hangul syllable blocks,
Devanagari conjuncts, or Maya glyph blocks --- and it offers a principled way to state what those
systems buy with the compression they give up.

\section{Limitations}
\label{sec:limits}

\paragraph{Stroke count is a proxy.}
It is the standard operationalisation of character complexity and is exactly measurable, but the
psycholinguistic literature indicates that recognition difficulty tracks overall complexity rather
than stroke count alone, and production cost surely depends on stroke length and curvature as well
as number. A component-count or ink-length operationalisation might shift the numbers, though the
strong monotone relation in Figure~\ref{fig:rank} is unlikely to be an artefact of the measure.

\paragraph{Character is not word.}
We treat the character as the coded unit, following \citet{deng2014rank}, but most modern Chinese
words are multi-character. A word-level analysis with word frequencies and summed stroke counts
would answer a related and arguably more comparable question; we do not attempt it here, and our
external comparison to word-length studies is correspondingly loose.

\paragraph{Corpus and standard dependence.}
Both frequency lists are large but each reflects a particular register and period, and the two
give $\Om$ values 6 points apart. Stroke counts follow the PRC standard; the Taiwan standard
differs for some components. The traditional-form condition covers only the 2{,}415
characters in the frequency list whose counts actually change under CC-CEDICT's single-character
mappings.

\paragraph{No causal or teleological claim.}
Our permutation test rejects chance arrangement; it does not identify a mechanism. Compression
pressure, phonological and semantic conservatism, and script-reform policy are all plausible
contributors, and the reform analysis is the only part of this paper with an identifiable agent.
We also make no claim that any of this was designed, other than the reform itself.

\paragraph{Truncation sensitivity.}
As shown in Section~\ref{sec:robust}, $\Om$ depends materially on inventory depth. Comparisons of
$\Om$ across studies should be treated as indicative until inventory depth is standardised.

\section{Conclusion}

Measured against its own achievable optimum, the simplified Chinese character inventory is
compressed to $\Om = 0.668$ --- two-thirds of the way from chance to the best possible assignment
of its existing shapes, replicated at 0.609 on an independent corpus of a different register, and
close to the 62--67\% band reported for word lengths in unrelated scripts. Because strokes come
from a closed five-element taxonomy, we can also say how far that is from optimal coding in
absolute terms: 7.22 strokes per token against a 5-ary Huffman optimum of 4.34, a factor of 1.66.
The residual gap is explained not by inefficiency but by architecture. With a Kraft sum of 2.05,
stroke sequences cannot be a uniquely decodable one-dimensional code; characters are distinguished
spatially, and the strokes that optimal coding would call redundant are what make the system
componential and partially transparent. The twentieth-century simplification reform, which raised
$\Om$ from 0.555 to 0.668 by shortening precisely the most frequent characters, shows that the
remaining slack was real and could be acted on deliberately --- but also that most of the
compression had already been achieved, slowly and without anyone intending it.

\subsection*{Data and code availability}
All analysis code and the derived per-character dataset (11{,}991 characters with stroke counts,
stroke sequences, frequencies and glosses) are available from the author on request. The analysis
uses no external dependencies beyond the Python standard library and \texttt{matplotlib}.

\subsection*{Acknowledgements}
This work uses the character-frequency list of Jun Da, the Leiden Weibo Corpus frequency data of
Daan van Esch, the CC-CEDICT dictionary, and a stroke-order database distributed with the
\texttt{chinese-stroke-sorting} package.

\end{document}